\documentclass[10pt,conference]{IEEEtran}
\usepackage{amssymb}
\usepackage{amsfonts,amsmath,graphicx,amssymb,algorithm, algorithmic,graphs}
\newcommand{\bl}{\boldsymbol}
\newcommand{\ml}{\mathcal}

\newtheorem{theorem}{Theorem}
\newtheorem{proposition}{Proposition}

\begin{document}

\title{Distributed Throughput-optimal Scheduling in\\ Ad Hoc Wireless Networks}
\author{\authorblockN{Qiao Li}
\authorblockA{qiaoli@cmu.edu\\
Department of Electrical and Computer Engineering \\
Carnegie Mellon University \\
5000 Forbes Ave., Pittsburgh, PA 15213} \and
\authorblockN{Rohit Negi}
\authorblockA{negi@ece.cmu.edu\\
Department of Electrical and Computer Engineering \\
Carnegie Mellon University \\
5000 Forbes Ave., Pittsburgh, PA 15213}}

\maketitle

\begin{abstract}
In this paper, we propose a distributed throughput-optimal ad hoc
wireless network scheduling algorithm, which is motivated by the
celebrated simplex algorithm for solving linear programming (LP)
problems. The scheduler stores a sparse set of basic schedules, and
chooses the max-weight basic schedule for transmission in each time
slot. At the same time, the scheduler tries to update the set of
basic schedules by searching for a new basic schedule in a
throughput increasing direction. We show that both of the above
procedures can be achieved in a distributed manner. Specifically, we
propose an average consensus based link contending algorithm to
implement the distributed max weight scheduling. Further, we show
that the basic schedule update can be implemented using CSMA
mechanisms, which is similar to the one proposed by Jiang \emph{et
al.} \cite{jiang10}. Compared to the optimal distributed scheduler
in \cite{jiang10}, where schedules change in a random walk fashion,
our algorithm has a better delay performance by achieving faster
schedule transitions in the steady state. The performance of the
algorithm is finally confirmed by simulation results.
\end{abstract}

\section{Introduction}
\label{sec_intro}

Scheduling in ad hoc wireless networks is, in general, an
NP-complete problem. In order to achieve throughput optimality, one
either needs to solve a problem with exponential complexity in each
time slot (e.g., the max-weight scheduler in \cite{tassiulas92}), or
by amortizing the complexity among an exponential number of time
slots (e.g., the random ``pick-and-compare'' scheduler in
\cite{tassiulas98}), so that the number of operations required in
each time slot is constant, at the expense of large (exponential in
the network size) queue lengths in the worst case \cite{shah10}.

\begin{figure}[t]
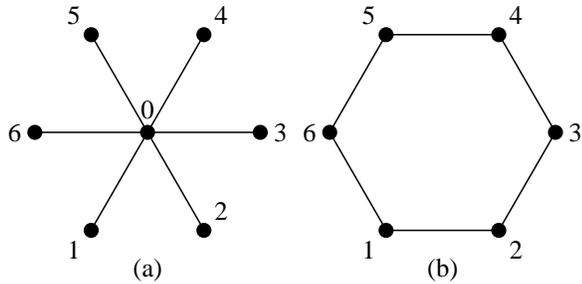

\begin{center}
\begin{tabular}{c c}
\begin{graph}(3.5,3.5)(0,0)
\roundnode{L0}(1.75, 1.5) \roundnode{L1}(1, 0.2) \roundnode{L2}(2.5,
0.2) \roundnode{L3}(3.25, 1.5) \roundnode{L4}(2.5, 2.8)
\roundnode{L5}(1, 2.8) \roundnode{L6}(0.25, 1.5)
\autonodetext{L0}[n]{0} \autonodetext{L1}[sw]{1}
\autonodetext{L2}[ne]{2} \autonodetext{L3}[e]{3}
\autonodetext{L4}[ne]{4} \autonodetext{L5}[nw]{5}
\autonodetext{L6}[w]{6} \edge{L0}{L1} \edge{L0}{L2} \edge{L0}{L3}
\edge{L0}{L4} \edge{L0}{L5} \edge{L0}{L6}
\end{graph}
&
\begin{graph}(3.5,3.5)(0,0)
\roundnode{L1}(1, 0.2) \roundnode{L2}(2.5, 0.2) \roundnode{L3}(3.25,
1.5) \roundnode{L4}(2.5, 2.8) \roundnode{L5}(1, 2.8)
\roundnode{L6}(0.25, 1.5) \edge{L1}{L2} \edge{L2}{L3} \edge{L3}{L4}
\edge{L4}{L5} \edge{L5}{L6} \edge{L6}{L1} \autonodetext{L1}[sw]{1}
\autonodetext{L2}[se]{2} \autonodetext{L3}[e]{3}
\autonodetext{L4}[ne]{4} \autonodetext{L5}[nw]{5}
\autonodetext{L6}[w]{6}
\end{graph}
\\
(a)&(b)
\end{tabular}
\end{center}
\caption{\label{fig:interference_graph}(a) is a star-shaped
interference graph of 7 links, and (b) is a ring-shaped interference
graph of 6 links.}
\end{figure}

In addition, the requirement that the scheduler be implemented in a
distributed manner makes the scheduling problem more difficult to
solve. In the literature, distributed scheduling algorithms are
often designed to be sub-optimal (e.g., the greedy schedulers in
\cite{chaporkar08}, \cite{li08}, \cite{li10}), so that minimal
coordinations among links are needed during scheduling. Recently,
Jiang \emph{et al.} \cite{jiang10} showed a surprising result that,
distributed throughput-optimal scheduling can be achieved by
cleverly adjusting the CSMA contending parameters, which is based on
the theory of Markov Chain Monte Carlo (MCMC). A throughput optimal
scheduler is one that can achieve the rate stability for any arrival
process whose average rate can be stabilized by \emph{some}
scheduler. This result has spurred interest among researchers in
searching for other distributed throughput-optimal scheduling
algorithms \cite{ni10}, \cite{shin09}, \cite{jiang10arxiv}. However,
since these current scheduling algorithms are all based on the MCMC,
they may suffer from large delays, even if they could begin in
steady state (i.e., with optimal CSMA parameters chosen initially).
This is due to the random walk behavior of the embedded time
reversible Markov chain, which requires a long time, in general, for
the schedule to change significantly. To illustrate this point,
consider the star-shaped interference graph in Fig.
\ref{fig:interference_graph} (a), where each node represents a link
in the wireless network, and each edge represents a transmission
conflict. Thus, the transmission schedules must be independent sets
of the graph, e.g., $\{0\}$, and $\{1,2,3,4,5,6\}$. Due to the
star-shaped topology, these two independent sets can also be viewed
as the two ``modes'' of this network. Now, the distributed CSMA
scheduling implies that the scheduled independent sets form a random
walk on the family of independent sets, where transitions are only
allowed between independent sets which differ by one link. For
example, the schedule can change from $\{1\}$ to $\{1,2\}$, but not
from $\{1\}$ to $\{0\}$, since that requires two simultaneous
changes: shutting down link 1 and activating link 0. Suppose the
current schedule is at mode $\{1,2,3,4,5,6\}$. In order to switch to
the other mode $\{0\}$, all of the links $\{1,2,3,4,5,6\}$ need to
stop transmission first, so that link 0 is able to contend for
transmission. Intuitively, this takes a long time, in particular, if
the other links are unwilling to stop transmission, due to their own
traffic demands. Thus, the distributed CSMA scheduling may incur a
large delay by spending a long time in each mode before transition,
since the random walk based design requires the scheduler to go
through the intermediate (sub-optimal) schedules before switching to
another mode, which happens with low probability. For more details
about the steady state delay, please see the queueing simulation
result in Fig. \ref{fig_ring_6}, which is explained in detail in
Section \ref{sec_simulation}.

Realizing the limitations of these distributed CSMA scheduling
algorithms, we try to improve the delay performance by avoiding the
random-walk based transitions between schedules. Instead of
allocating positive probabilities to all the schedules, as required
by MCMC, we require that the scheduler choose the transmitting
schedules among \emph{a sparse set of explicitly enumerated
``modes'' (basic schedules)}, so that the schedule transitions are
faster, since there is no need for the scheduler to enter the
sub-optimal ``transition modes''. Further, since the number of basic
schedules is small, optimal scheduling, \emph{from among these
schedules}, can be achieved efficiently in a distributed fashion.
This is done by solving a simple max-weight independent set problem,
which can be achieved by a certain consensus algorithm.

The existence of such a sparse optimal solution is guaranteed by the
well-known Carath\'eodory theorem \cite{bertsekas03}, which states
that, at most $n+1$ schedules are needed to achieve the optimal
scheduling in a network with $n$ links. Further, such a solution can
be efficiently computed by the celebrated simplex algorithm
\cite{bertsimas97}, which can be viewed as a steepest descent
algorithm along the edges of the feasible region (a polytope for
LP). Inspired by the simplex algorithm, we propose a distributed
algorithm to search for the optimal basic schedules, which can be
implemented using distributed CSMA mechanisms similar to
\cite{jiang10}. Thus, by combining the consensus based scheduling
and CSMA based basic schedule updates, we show that throughput
optimal scheduling can be achieved in a distributed fashion.
Finally, the delay improvement is verified in the simulation
section.

The organization of the rest of the paper is as follows: In Section
\ref{sec_system} we introduce the system model, and in Section
\ref{sec_algorithm} we propose and analyze the scheduling algorithm.
Section \ref{sec_simulation} shows the simulation results, and
finally Section \ref{sec_conclusion} concludes this paper.

\section{System Model}
\label{sec_system}

In this section we introduce the system model, which is standard in
the literature. We first introduce the network model.

\subsection{Network Model}

We are interested in the scheduling problem at the MAC layer of a
wireless network, where the topology of the network is described by
an interference graph $\ml G_I=(\ml V_I, \ml E_I)$, where $\ml V_I$
is the set of $n$ links, and $\ml E_I$ is the set of pairwise
interference constraints, i.e., link $i$ and link $j$ are not
allowed to transmit together if $(i,j)\in \ml E_I$, due to the
strong interference that one link causes upon the other. We assume a
slotted time system, and in each time slot $t$, the scheduled
transmitting links $\bl \sigma_{\text{sch}}(t)$ must form an
independent set in $\ml G_I$. With an abuse of notation, we also
denote $\bl \sigma_{\text{sch}}(t)$ as an $n\times 1$ vector, where
$\sigma_{\text{sch}, i}(t)=1$ if link $i$ belongs to the
transmitting independent set, and $\sigma_{\text{sch}, i}(t)=0$
otherwise.

The queueing dynamics of the network is as follows:
\begin{eqnarray}
\bl Q(t)=\bl Q(0)+\bl A(t)-\bl D(t)
\end{eqnarray}
where $\bl Q(t)$ is the queue length vector at time slot $t$, and
$\bl A(t)$ and $\bl D(t)$ are the number of cumulative arrived and
departed packets during the first $t$ time slots, respectively. We
assume that the packet arrival process is subject to the Strong Law
of Large Numbers (SLLN), i.e., with probability 1 (w.p.1), we have
$\lim_{t\rightarrow\infty} \bl A(t)/t=\bl a$ where $\bl a$ is the
arrival rate vector. Further, we also assume that the number of
arrived packets in each time slot is uniformly bounded by a large
constant. Note that this assumption is quite mild, since the packet
arrivals are allowed to be correlated both across time slots, and
different links. Thus, our model is well suited in analyzing typical
wireless networks, and can be readily generalized to multi-hop
scenarios.

A basic requirement on the scheduler is the \emph{rate stability},
i.e., for each link $i$, we have $\lim_{t\rightarrow\infty}
D_i(t)/t=a_i$, w.p.1, so that an average throughput of $\bl a$ can
be achieved by the scheduler. Thus, a \emph{throughput optimal}
scheduler should achieve rate stability for any $\bl a$ which can be
stabilized by \emph{some} scheduler. In the seminal paper
\cite{tassiulas92}, Tassiulas \emph{et al.} showed that the optimal
stability region is $\ml A={\bf Co}(\ml M)$, where ${\bf Co}(\cdot)$
denotes the convex hull, and $\ml M$ is the family of all
independent sets. Further, they propose the following optimal
max-weight scheduling algorithm:
\begin{equation}\label{eqn_mw_tassiulas}
\bl\sigma_{\text{sch}}(t)=\arg\max_{\bl\sigma\in \ml
M}\bl\sigma^T\bl Q(t)
\end{equation}
However, such an algorithm requires solving an NP-complete problem
in every time slot, and is very difficult to implement in a
distributed manner. In the next subsection we introduce the recently
developed throughput optimal distributed CSMA scheduling algorithm.

\subsection{Throughput Optimal Distributed Scheduling}

Before introducing the distributed CSMA scheduling algorithm, we
need to have an optimization interpretation of the scheduling
problem. We can formulate the scheduling problem as the following
feasibility problem:
\begin{eqnarray}
\textsf{SCH: }\text{minimize}_{\bl x} &&0\\
\text{subject to}&&M\bl x=\bl a\label{eqn_rate_constraint}\\
&& \bl 1^T\bl x=1, \bl x\succeq \bl 0\label{eqn_simplex_constraint}
\end{eqnarray}
where $\bl a$ is the arrival rate vector, $M$ is the matrix whose
columns are all the independent sets, and $\bl x$ is the scheduling
variable, such that $x_{\bl \sigma}$ represents the asymptotic time
fraction that independent set $\bl \sigma$ (which is a column of the
matrix $M$) is chosen by the scheduler. Thus, $\bl x$ naturally
lives in the simplex, as described by
(\ref{eqn_simplex_constraint}), and (\ref{eqn_rate_constraint}) is
essentially the rate stability constraint.

Note that \textsf{SCH} has many solutions, due to the key structure
in the highly under-determined system in
(\ref{eqn_simplex_constraint}) that, any subset of an independent
set is also an independent set. Thus, based on a solution for any
strictly feasible arrival rate vector $\bl a$, one can easily
construct a set of feasible solutions. However, distributed
implementation of any solution is a very challenging problem. In
\cite{jiang10}, Jiang \emph{et al.} obtains a solution in the
exponential family by transforming \textsf{SCH} into the following
max-entropy problem \textsf{ME}:
\begin{eqnarray*}
\textsf{ME: }\text{maximize}_{\bl x} &&H(\bl x)=-\sum_{\bl \sigma\in\ml M}x_{\bl \sigma}\log x_{\bl \sigma}\\
\text{subject to}&& (\ref{eqn_rate_constraint})\text{ and }
(\ref{eqn_simplex_constraint})
\end{eqnarray*}
It can be shown that the solution $\bl x^\star$ has the following
form:
\begin{equation}\label{eqn_x_exp}
x_{\bl \sigma}^\star = \exp(\bl \theta^{\star T}\bl \sigma-A(\bl
\theta^\star))
\end{equation}
where $\bl \theta^\star$ corresponds to the Lagrange multiplier
associated with the rate constraints in (\ref{eqn_rate_constraint}),
and
\begin{equation}
A(\bl\theta^\star)=\log\sum_{\bl \sigma\in\ml
M}\exp(\bl\theta^{\star T}\bl \sigma)
\end{equation}
is the standard log-partition function. Following the literature of
Gibbs sampling (an example of MCMC), it was shown in \cite{jiang10}
that the optimal allocations $\bl x^\star$ can be implemented by a
time-reversible Markov chain, using CSMA mechanisms. Further, the
optimal parameters (Lagrange multipliers) $\bl \theta^\star$ can be
obtained using a stochastic gradient algorithm. In below we describe
a discrete-time version of the distributed scheduling algorithm,
which is also closely related to the algorithm in \cite{ni10}.

\begin{algorithm}
\caption{\textsf{CSMA}($\bl \theta$)} \label{alg_csma}
\begin{algorithmic}
\STATE In each time slot $t$, do the following:

\STATE Randomly generate an independent set $\bl \sigma(t)$ from a
certain distribution, independently across time slots.

\FOR {each $i\in\bl \sigma(t)$}

\STATE $p_i=\exp(\theta_i)/(1+\exp(\theta_i))$;

\IF{ no neighbor of $i$ is in $\bl \sigma_{\text{csma}}(t-1)$}

\STATE
\begin{equation*}
\sigma_{\text{csma}, i}(t) = \left\{ \begin{array}{ll}
         1 & \textrm{ with probability } p_i\\
                  0& \textrm{ else }
                  \end{array} \right.
\end{equation*}

\ENDIF

\ENDFOR

\STATE Any other link $i$ not in $\bl \sigma(t)$ set $
\sigma_{\text{csma}, i}(t)=\sigma_{\text{csma}, i}(t-1)$.

\end{algorithmic}
\end{algorithm}

In the above algorithm, carrier sensing is used in two phases: 1)
generation of the independent set $\bl\sigma(t)$ (according to
certain protocol \cite{ni10}) and 2) detection of whether there is a
transmitting neighbor of a link $i\in\bl\sigma(t)$ at time slot
$t-1$. Further, note that the key part of the algorithm, phase 2, is
fully distributed, with no explicit message exchange among links.
The following proposition shows that the above algorithm achieves
the distribution in (\ref{eqn_x_exp}) asymptotically.
\begin{proposition}
$\bl \sigma_{\text{csma}}(t)$ in \textsf{CSMA}$(\bl\theta^\star)$
form a time-reversible Markov chain, with the steady state
distribution in the form of (\ref{eqn_x_exp}).
\end{proposition}
\begin{IEEEproof}
The claim is proved by checking balance equations. The proof is
essentially the same as the one in \cite{ni10}, which we omit due to
space limitation.
\end{IEEEproof}

As discussed in Section \ref{sec_intro}, however, the above
scheduling algorithm suffers from long delay, due to the random walk
behavior associated with the time-reversible Markov chain. In the
following section we try to solve this problem by introducing the
simplex scheduling algorithm.

\section{Simplex Scheduling}
\label{sec_algorithm}

In this section we propose optimal distributed scheduling, which is
based on the simplex algorithm. For a detailed description of the
simplex algorithm for general LP problems, please see, for example,
\cite{bertsimas97}. Since the simplex algorithm requires a feasible
starting point, we next transform the problem \textsf{SCH} into a
relaxed problem \textsf{REL}, where a feasible initial vertex is
easy to obtain:
\begin{eqnarray}
\textsf{REL: }\text{minimize}_{\bl x, \gamma} &&\gamma\\
\text{subject to}&&M\bl x=(1-\gamma)\bl a\\
&& \bl 1^T\bl x=1, \bl x\succeq \bl 0, 0\leq\gamma\leq 1
\end{eqnarray}
In above, the relaxation variable $\gamma$ can be interpreted as the
``throughput gap'', so that $\gamma^\star=0$ if and only if the
arrival rate $\bl a$ can be stabilized by the scheduler. In order to
illustrate the main ideas of the distributed scheduling algorithm,
we first introduce a hypothesized centralized simplex algorithm to
solve \textsf{REL}.

\subsection{Centralized Simplex Scheduling}

The centralized simplex algorithm \textsf{SIM} is shown in Algorithm
\ref{alg_simplex}, which is essentially an application of the
general simplex algorithm to the specific scheduling problem
\textsf{REL}.
\begin{algorithm}
\caption{\textsf{SIM}} \label{alg_simplex}
\begin{algorithmic}
\STATE {\bf Initialization}:

$B=I, \bl x_B^\star = \min({1\over \bl 1^T \bl a}, 1)\bl a,
\gamma^\star=\max(1-{1\over \bl 1^T \bl a}, 0)$.

\WHILE{$\gamma^\star>0$}

\STATE
\begin{enumerate}
\item {\bf Simplex Search}: Compute the moving direction
\begin{eqnarray}\label{eqn_mwis}
\bl \sigma_{\text{new}} = \arg\max_{\bl \sigma\in M}\bl 1^TB^{-1}\bl
\sigma\
\end{eqnarray}
\item {\bf Scheduling}: Compute the new vertex $(\bl x_B^\star, x_{\text{new}}^\star)$ and the throughput gap $\gamma^\star$ by solving
\begin{eqnarray*}
\textsf{MOV: }\min_{\bl x_B, x_{\text{new}}, \gamma}&& \gamma\\
\text{subject to}&& B\bl x_B+\bl \sigma_{\text{new}} x_{\text{new}} =(1-\gamma)\bl a\\
&&\bl 1^T\bl x_B+x_{\text{new}}=1\\
&&\bl x_B\succeq \bl 0, x_{\text{new}}\geq 0, 0\leq\gamma\leq 1
\end{eqnarray*}
\item {\bf Update}: Let be $\bl \sigma$ be a column in
$B$ such that $x^\star_{\bl \sigma}=0$. Replace $\bl \sigma$ with
$\bl \sigma_{\text{new}}$, and relabel the variables.
\end{enumerate}

\ENDWHILE

\RETURN $(B, \bl x_B^\star, \bl \gamma^\star)$
\end{algorithmic}
\end{algorithm}

The above simplex algorithm solves the problem \textsf{REL} by
moving along adjacent vertices of the feasible region in a cost
reducing direction. In order to understand its behavior, we first
need to identify a vertex. According to the equality constraints in
\textsf{REL}, a vertex (or a \emph{basic solution} in the simplex
algorithm terminology) $(\bl x_B, \gamma)$ can be uniquely
determined by a basis matrix as follows:
\begin{equation}\label{eqn_null_space}
\left(
\begin{matrix}
B & \bl a\\
\bl 1^T & 0
\end{matrix}\right)
\left(
\begin{matrix}
\bl x_B \\ \gamma
\end{matrix}\right)=\left(
\begin{matrix}
\bl a\\ 1
\end{matrix}\right)
\end{equation}
where $B$ is an $n\times n$ (invertible) sub-matrix of $M$, which
represents $n$ independent sets, and $\bl x_B$ is an $n\times 1$
sub-vector of $\bl x$, which are the allocated time fractions of the
independent sets in $B$. It is easy to verify that the initial
vertex as specified in the initialization phase of \textsf{SIM} is
feasible.

We next show that \textsf{SIM} successfully moves to an adjacent
vertex following a cost reducing direction. An edge is represented
by a new column vector $\bl \sigma\choose\bl 1$, where $\bl \sigma$
is an independent set, and moving along the edge is equivalent to
increasing the coefficient associated with $\bl \sigma\choose\bl 1$,
with the constraint that the equalities in \textsf{REL} still hold.
Thus, the changes of the optimization variables must stay in the
null space of the matrix $\left(
\begin{matrix}
M & \bl a\\
\bl 1^T & 0
\end{matrix}\right)$, i.e., we increase the coefficient of $\bl \sigma
\choose 1$ by a unit, the changes in the existing variables are
\begin{equation}\left(
\begin{matrix}
B & \bl a\\
\bl 1^T & 0
\end{matrix}\right)
\left(
\begin{matrix}
\Delta\bl x_B \\ \Delta\gamma
\end{matrix}\right)+\left(
\begin{matrix}
\bl \sigma \\1
\end{matrix}\right)=\bl 0
\end{equation}
Using the block matrix inversion formula, we obtain the rate of
change in the cost:
\begin{eqnarray}
\Delta\gamma = (1-\gamma)(1-\bl 1^TB^{-1}\bl \sigma)
\end{eqnarray}
Thus, we need to find a proper independent set $\bl\sigma$, so that
the cost change $\Delta\gamma\leq 0$. We next show that the $\bl
\sigma^\star$ obtained in (\ref{eqn_mwis}) is a proper cost-reducing
direction.

\begin{proposition}\label{prop_cost_reducing}
The cost change for the independent set $\bl \sigma^\star$ in
(\ref{eqn_mwis}) satisfies $\Delta\gamma\leq 0$, and the inequality
is strict if $\gamma>0$.
\end{proposition}
\begin{IEEEproof}
From (\ref{eqn_null_space}) we have $\bl x_B = (1-\gamma)B^{-1}\bl
a$. After multiplying both sides with $\bl 1^T$ and noting that $\bl
1^T\bl x_B=1$, we have
\begin{equation}
(1-\gamma)(\bl 1^TB^{-1}\bl a)=1
\end{equation}
Noting that $\bl a\in {\bf Co}(\ml M)$ and (\ref{eqn_mwis}), we have
\begin{equation}
\bl 1^TB^{-1}\bl \sigma^\star\geq\bl 1^TB^{-1}\bl a
\end{equation}
and therefore
\begin{eqnarray*}
\Delta\gamma&=&(1-\gamma)(1-\bl 1^TB^{-1}\bl \sigma^\star)\\
&\leq& (1-\gamma)(1-\bl 1^TB^{-1}\bl a)=-\gamma
\end{eqnarray*}
from which the claim follows.
\end{IEEEproof}

Given the new direction specified by $\bl \sigma^\star$, according
to the standard simplex algorithm, \textsf{SIM} then moves along the
edge as specified by ${\bl \sigma^\star\choose 1}$, until it reaches
a new vertex, where some coefficient of the independent set in $B$
first becomes zero. Then \textsf{SIM} replaces that column with $\bl
\sigma^\star$, and relabel the variables if necessary. We next show
that, this movement to a better vertex is achieved by solving the
problem \textsf{MOV}.
\begin{proposition}
For the solution $(\bl x_B^\star, x_{\text{new}}^\star,
\gamma^\star)$ to the problem \textsf{MOV}, we have
$x_{\text{new}}^\star>0$, and there is one column $\bl \sigma$ in
$B$ such that $x^\star_{\bl \sigma}=0$.
\end{proposition}
\begin{IEEEproof}
Due to space limitation, we only describe the intuition behind the
proof. Suppose $x_{{\text{new}}}^\star=0$, then the solution to
\textsf{MOV} is at the same vertex associated with the old matrix
$B$, which contradicts the fact that $\bl \sigma_{\text{new}}$ is a
cost-reducing direction, according to Proposition
\ref{prop_cost_reducing}. The claim that some $\bl \sigma$ in $B$
has $x^\star_{\bl \sigma}=0$ follows from the fact that \textsf{MOV}
has bounded optimum.
\end{IEEEproof}

Having shown that the centralized algorithm \textsf{SIM} is the same
as the general simplex algorithm for LP problems, we have the
following conclusion:
\begin{theorem}\label{theorem_simplex}
If $\bl a\in{\bf Co}(\ml M)$, \textsf{SIM} will return a solution
$(B, \bl x_B^\star, \gamma^\star)$ such that $\gamma^\star=0$, and
$(\bl x_B^\star, \gamma^\star)$ solves \textsf{REL}.
\end{theorem}

In the next subsection we show that this algorithm can be
efficiently implemented in a distributed manner, using CSMA and
average consensus algorithms.

\subsection{Distributed Simplex Scheduling}

In this subsection we propose the distributed scheduling problem,
and prove its throughput optimality, by showing that it is a
distributed implementation of the centralized simplex scheduling
algorithm in the last subsection. We first illustrate the full
scheduling algorithm \textsf{DIS-SCH} in Algorithm
\ref{alg_scheduling}.

\begin{algorithm}
\caption{\textsf{DIS-SCH}} \label{alg_scheduling}
\begin{algorithmic}
\STATE {\bf Initialization}: Set the initial parameters as $B(0)=I$,
$\bl \theta(0)=\bl 0$, $\gamma(0) = 1$, $\bl \sigma_{\text{csma}}(0)
= \bl 0$, and $\bl \sigma_{\text{new}}(0)=\bl 0$.

\STATE In each time slot $t$, do the following:
\begin{enumerate}
\item {\bf CSMA}: update $\bl \sigma_{\text{csma}}(t)$ by running
\textsf{CSMA}($\alpha\bl \theta(t)$), where $\alpha$ is a
sufficiently large constant.
\item {\bf Scheduling}: for each link $i$, compute
\begin{equation}\label{eqn_simple_mwis}
\sigma_{\text{sch}, i}(t)=\arg\max_{\bl \sigma\in B\cup\{\bl
\sigma_{\text{new}}\}} w^{(i)}_{\bl \sigma}(t)
\end{equation}
Link $i$ transmits if $\sigma_{\text{sch}, i}(t)=1$. In above,
$w_{\bl \sigma}(t)=\bl\theta(t)^T\bl \sigma$ is the weight of
independent set $\bl \sigma$, and $w^{(i)}_{\bl \sigma}(t)$ is link
$i$'s local copy, which is updated by the consensus algorithm.

\item {\bf Consensus}: update the parameters
\begin{eqnarray}
\bl \theta(t) &=& \bl \theta(t)+\epsilon(\bl (1-\gamma(t))\bl
a(t)-\bl \sigma_{\text{sch}}(t))\label{eqn_theta_update}\\
\gamma(t) &=& [\gamma(t)+\epsilon(\bl \theta(t)^T\bl
a(t)-1)]_{[0,1]}\label{eqn_gamma_update}\\
\bl a(t) &=&\bl A(t)/t
\end{eqnarray}
where $[\cdot]_{[0,1]}$ means projecting onto the interval $[0,1]$,
and $\epsilon$ is a constant step size. Run an average consensus
algorithm over the quantities $\{w_{\bl \sigma}(t)\}_{\bl \sigma\in
B\cup\{\bl \sigma_{\text{new}}\}}$ and $\gamma(t)$.
\item {\bf Update}: If certain convergence conditions are satisfied,
replace the column in $B$ with the smallest weight by $\bl
\sigma_{\text{new}}(t)$, and load the CSMA independent set by
letting $\bl\sigma_{\text{new}}(t)=\bl\sigma_{\text{csma}}(t)$.
\end{enumerate}

\end{algorithmic}
\end{algorithm}

Overall \textsf{DIS-SCH} is very similar to \textsf{SIM}, with the
following two important changes: 1) the simplex search phase in
\textsf{SIM} is replaced with the CSMA phase, and that 2) the
centralized moving procedure \textsf{MOV} is replaced with a
\emph{simple max-weight scheduling} in (\ref{eqn_simple_mwis}),
which is then followed by parameter updates and average consensus
algorithm. It may appear that (\ref{eqn_simple_mwis}) is similar to
the max-weight scheduler in (\ref{eqn_mw_tassiulas}), with the
difference that the queue lengths $\bl Q(t)$ are replaced by the
``virtual queue lengths'' $\bl\theta(t)$. However, the max-weight
independent set in (\ref{eqn_simple_mwis}) is chosen from $n+1$
independent sets, \emph{instead of the family of all independent
sets, which has exponential size}. Thus, (\ref{eqn_simple_mwis}) can
be solved efficiently (in linear time), whereas the max-weight
scheduler in (\ref{eqn_mw_tassiulas}) is NP-complete, in general. In
the following we elaborate on the change in Step 2, by showing that
the scheduling and consensus phases in \textsf{DIS-SCH} is
equivalent to solving the problem \textsf{MOV}.

\begin{proposition}
If there is no change in the matrix $B$, the parameters $\bl \theta$
updated in (\ref{eqn_theta_update}) will converge to the optimal
dual variable for \textsf{MOV}, $\gamma$ will converge to the
optimal cost $\gamma^\star$, and the average time fractions $(\bl
x_B, x_{{\text{new}}})$ will converge to the optimal primal
variables $(\bl x^\star_B, x^\star_{{\text{new}}})$ for
\textsf{MOV}.
\end{proposition}

\begin{IEEEproof}
Due to space limits, we only describe the intuition behind the
proof. We first form the Lagrangian of \textsf{MOV}
\begin{eqnarray*}
\text{minimize}_{\bl x_B, x_{\text{new}}, \gamma} &&\gamma+\bl \theta^T((1-\gamma)\bl a-B\bl x_B-\bl \sigma_{\text{new}} x_{\text{new}})\\
\text{subject to}&& \bl 1^T\bl x_B+x_{\text{new}}=1, \bl x_B\succeq
\bl 0, x_{\text{new}}\geq 0
\end{eqnarray*}
Note that this is a LP over a simplex, and therefore, the solution
is obtained at a vertex. Specifically, we can choose the vertex
$x_{\bl \sigma_{\text{sch}}}=1$ (which is equivalent to scheduling
independent set $\bl \sigma_{\text{sch}}$), where $\bl
\sigma_{\text{sch}}$ satisfies
\begin{equation}\label{eqn_simple_mwis_centralized}
\bl \sigma_{\text{sch}}\in\arg\max_{\bl \sigma\in B\cup\{\bl
\sigma_{\text{new}}\}}\bl \theta^T\bl \sigma
\end{equation}
Further, under mild assumptions about the consensus algorithm, it
can be shown that the two schedules in (\ref{eqn_simple_mwis}) and
(\ref{eqn_simple_mwis_centralized}) are the same with high
probability. Finally, it can be shown that the updates in
(\ref{eqn_theta_update}) and (\ref{eqn_gamma_update}) correspond to
the standard primal-dual algorithms for solving convex optimization
problems. Thus, the variables will converge to the optimal.
\end{IEEEproof}

We next elaborate on the change in Step 1, by showing the
equivalence between the CSMA phase in \textsf{DIS-SCH} and the
simplex search phase in \textsf{SIM}. We have the following
proposition.

\begin{proposition}
With sufficiently large $\alpha$ and sufficiently long time, $\bl
\sigma_{\text{csma}}(t)$ is the max-weight independent set in
(\ref{eqn_mwis}), with high probability.
\end{proposition}
\begin{IEEEproof}
Due to space limits, we only describe the intuition behind the
proof. From the complementary slackness property, from the
Lagrangian of \textsf{MOV} we have
\begin{equation}
\bl \theta^{\star T}B=(1-\gamma^\star)(\bl \theta^\star\bl a)\bl 1^T
\end{equation}
Note that for notation simplicity, $B$ has already been updated by
replacing one sub-optimal column with $\bl\sigma_{\text{new}}$,
i.e., $B=B_{\text{new}}$. Thus, the optimal dual variables $\bl
\theta^\star$ satisfies $\bl \theta^\star \propto\bl 1^T B^{-1}$,
and therefore, if $\bl \sigma_{\text{csma}}(t)$ satisfies
\begin{equation}
\bl \sigma_{\text{csma}}(t) \in \arg\max_{\bl \sigma \in M}\bl
\theta^{\star T}\bl \sigma
\end{equation}
then it is essentially the same max-weight solution of
(\ref{eqn_mwis}). Further, note that the CSMA based sampling phase
converges slowly (exponential time, in general), whereas solving
\textsf{MOV} are relatively fast. Thus, we can assume that the
parameters $\bl \theta(t)$ have converged to the optimal, and
finally, the claim follows from the steady state distribution in
(\ref{eqn_x_exp}), and the well known approximation that
\begin{equation}
\exp(\alpha\bl \theta^{\star T}\bl
\sigma-A(\alpha\bl\theta^\star))\approx{\bl
1}_{\{\bl\sigma\in\arg\max_{\bl \sigma \in M}\bl \theta^{\star T}\bl
\sigma\}}
\end{equation}
for large enough $\alpha$, where ${\bl 1}_{\{\cdot\}}$ is the
indicator function, i.e., ${\bl 1}_{\{\text{true}\}}=1$ and ${\bl
1}_{\{\text{false}\}}=0$.
\end{IEEEproof}

Having shown that the distributed algorithm \textsf{DIS-SCH} is
essentially the same as the centralized simplex algorithm
\textsf{SIM}, we have the following conclusion:
\begin{theorem}\label{theorem_distributed}
If $\bl a\in\ml A$, \textsf{DIS-SCH} will return a solution $(B, \bl
x_B^\star, \gamma^\star)$ such that $\gamma^\star=0$, and $(\bl
x_B^\star, \gamma^\star)$ solves \textsf{REL}.
\end{theorem}

In the next section we will demonstrate the performance of
\textsf{DIS-SCH} by simulation results.

\section{Simulation Results}
\label{sec_simulation}

We next compare the performance of the distributed simplex
scheduling algorithm \textsf{DIS-SCH} with the distributed CSMA
scheduling algorithm in Algorithm \ref{alg_csma} using MATLAB
simulation. During the simulation, we assume that the packet
arrivals are i.i.d, with $95\%$ of the maximum uniform arrival rate.
The average consensus algorithm for the distributed simplex
scheduling is implemented as follows: We assume that the
communication graph for the average consensus is the same as the
interference graph. In each time slot, a random maximal matching is
first formed, and each matched pair update their variables by taking
the average of their local copies. We also allow a certain time
period for the average consensus algorithm to converge, before it is
used for distributed scheduling.

\subsection{Star Network}

\begin{figure}
\centering
\includegraphics[width=3.7in]{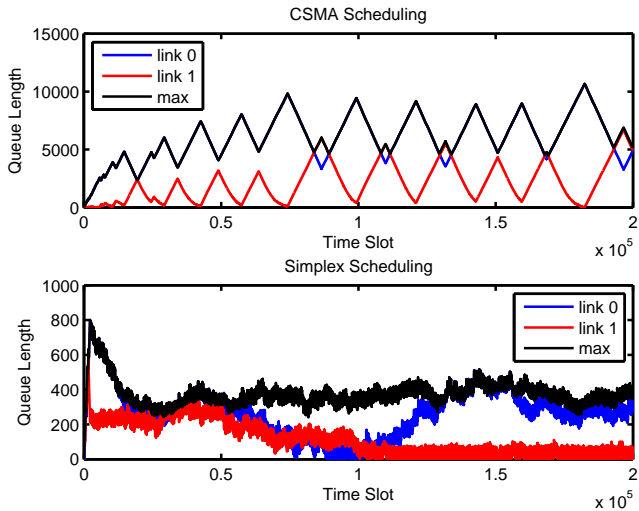}
\caption{The simulation result of a 7-star network with distributed
CSMA scheduling and simplex scheduling.} \label{fig_star_7}
\end{figure}

We fist consider the 7-link star-shaped interference graph in Fig
\ref{fig:interference_graph} (a), with the simulation result shown
in Fig. \ref{fig_star_7}. In the figure, the sample queue lengths
are plotted  over a simulation period of $2\times10^5$ time slots.
From the figure, one can observe that the network is rate stable in
both cases, but distributed CSMA scheduling has much larger queue
lengths (around $10^4$) than simplex scheduling (several hundreds)
in the steady state. Further, one can observe that link 0 is the
bottle neck link for CSMA scheduling, since its queue length is most
often the largest. This is because, as discussed in Section
\ref{sec_intro}, in the steady state, distributed CSMA scheduling
spends a considerable amount of time around each mode before
transiting to the intermediate (suboptimal) independent sets. Thus,
the transitions of CSMA scheduling is very slow, and the queue
lengths are consequently large. On the other hand, simplex
scheduling can quickly switch between the two optimal modes, and
therefore, have smaller queue lengths in the steady state.

\subsection{Ring Network}

\begin{figure}
\centering
\includegraphics[width=3.7in]{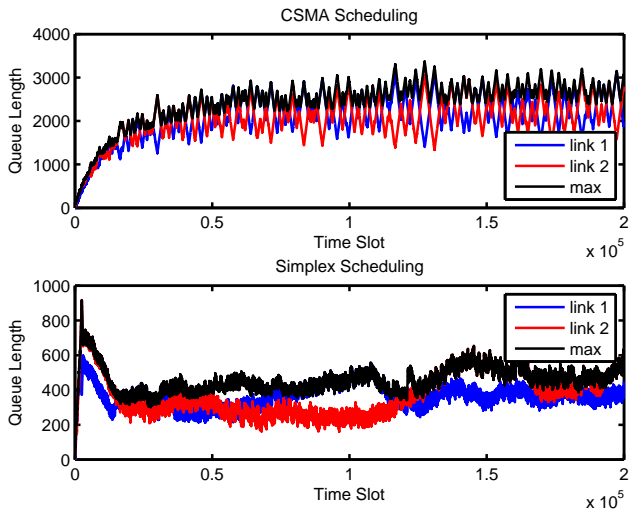}
\caption{The simulation result of a 6-ring network with different
distributed CSMA scheduling and simplex scheduling.}
\label{fig_ring_6}
\end{figure}

We next consider the 6-link ring-shaped interference graph in Fig
\ref{fig:interference_graph} (b). The simulation result is shown in
Fig. \ref{fig_ring_6}. Similar to the star network, one can observe
that both algorithm result in rate stability, but the simplex
scheduling achieves much smaller queue lengths than the distributed
CSMA scheduling. In fact, for the ring shaped network, it is easy to
see that the optimal modes are $\bl\sigma_1=(1, 0, 1, 0, 1, 0)^T$
and $\bl\sigma_2=(0, 1, 0, 1, 0, 1)^T$. In the steady state, the
simplex scheduler can achieve low delay by quickly switching between
these two modes, using the simple max-weight scheduler implemented
by average consensus. On the other hand, the switching is much more
time-consuming for the distributed CSMA scheduling, due to the
random walk based design. Finally, note that the transition time and
queue lengths for distributed CSMA scheduling in the 6-ring network
is smaller than that for the 7-star network (both have the same
arrival rates), due to the fact that, it is easier to switch between
the modes in a ring-shaped topology.

\section{Conclusion}
\label{sec_conclusion}

In this paper, we proposed a distributed throughput-optimal
scheduling algorithm for ad hoc wireless networks, which is
motivated by the simplex algorithm for solving LP problems. The
scheduler maintains a sparse set of basic schedules, and during
scheduling, the basic schedule with the maximum weight is selected
for transmission in each time slot. The set of basic schedules are
updated according to the simplex algorithm, which can be implemented
using CSMA mechanisms in a distributed fashion. Compared to the
distributed CSMA based scheduling in \cite{jiang10}, our algorithm
achieves better delay performance in the steady state by allowing
faster transitions among the optimal independent sets.

\section*{Acknowledgement}

The authors would like to thank the anonymous reviewers for their
constructive suggestions and comments. This work was supported in
part by the US NSF awards CNS-0831973 and ECCS-0931978, and by US
ARO award W911NF0710287.

\end{document}